%% file: 0-main.tex
\documentclass[sigconf]{acmart}

\copyrightyear{2026}
\acmYear{2026}
\setcopyright{cc}
\setcctype{by}
\acmConference[ICSE-Companion '26]{2026 IEEE/ACM 48th International Conference on Software Engineering}{April 12--18, 2026}{Rio de Janeiro, Brazil}
\acmBooktitle{2026 IEEE/ACM 48th International Conference on Software Engineering (ICSE-Companion '26), April 12--18, 2026, Rio de Janeiro, Brazil}
\acmPrice{}
\acmDOI{10.1145/3774748.3787621}
\acmISBN{979-8-4007-2296-7/2026/04}

\AtBeginDocument{%
  }

\usepackage{tikz}
\newcommand*\circled[1]{\tikz[baseline=(char.base)]{
    \node[shape=circle,draw,inner sep=0.5pt] (char) {\small#1};}}

\begin{document}
\title{Data-driven Test Generation for Fuzzing AI Compiler}

\author{Qingchao Shen}
\affiliation{
    \institution{School of Computer Software, Tianjin University}
    \city{Tianjin}
    \country{China}
}
\email{qingchao@tju.edu.cn}

\begin{abstract}
Artificial Intelligence (AI) compilers are critical for efficiently deploying AI models across diverse hardware platforms. However, they remain prone to bugs that can compromise both compiler reliability and model correctness. 
Thus, ensuring the quality of AI compilers is crucial.
In this work, we present a unified data-driven testing framework that systematically addresses stage-specific challenges in AI compilers. 
Specifically, OPERA migrates tests for AI libraries to test various operator conversion logic in the model loading stage. OATest synthesizes diverse optimization-aware computational graphs for testing high-level optimizations. HARMONY generates and mutates diverse low-level IR seeds to generate hardware-optimization-aware tests for testing low-level optimizations. Together, these techniques provide a comprehensive, stage-aware framework that enhances testing coverage and effectiveness, detecting 266 previously unknown bugs in four widely used AI compilers.

\end{abstract}

\maketitle

\input{1-intro}
\input{3-related}
\input{2-approach}
\input{4-conclusion}

\bibliographystyle{ACM-Reference-Format}
\bibliography{reference}

\end{document}

%% file: 1-intro.tex
\section{Introduction}
\label{intro}

Artificial Intelligence (AI) compilers (e.g., TVM~\cite{TVM}, TensorRT~\cite{TensorRT}, ONNXRuntime~\cite{onnx}) enable efficient deployment of AI models by performing a series of optimizations to generate efficient executable code for diverse hardware (e.g., GPUs~\cite{gpu} and TPUs~\cite{tpu}).
The compilation process of an AI model typically consists of three main stages~\cite{DLCStudy}: \circled{1} Loading the AI model, developed with a specific AI libraries (e.g., PyTorch~\cite{pytorch} and Keras~\cite{keras}), into an equivalent high-level intermediate representation (IR); \circled{2} Applying hardware-independent optimizations on the high-level IR; and \circled{3} Lowering the high-level IR into a low-level IR followed by hardware-specific optimizations to generate executable code for the target hardware.

Like other software~\cite{qaqa,ma2025bounded,desil,mlirsmith}, AI compilers are also prone to bugs, which may lead to severe consequences in both the compilers themselves and the models they produce. 
Recent studies~\cite{DLCStudy,dlc_evolution,huang2025false} reported that bugs are prevalent across all compilation stages, highlighting the urgent need for comprehensive testing to ensure the reliability of AI compilers.

\textbf{Challenges:}
Recently, several techniques have been proposed for testing AI compilers, including NNSmith~\cite{NNSmith}, HirGen~\cite{HirGen}, WhiteFox~\cite{whitefox}, Tzer~\cite{tzer}, and TVMFuzz~\cite{TVMFuzz}.
These approaches primarily focus on generating valid and diverse tests across different input formats, including AI models, high-level IRs, and low-level IRs.
Despite the progress, testing AI compilers still poses three fundamental challenges:
(1) \textit{Diverse frontend coverage:}
The model loading stage must correctly convert operators from various AI libraries, covering different operator types (e.g., Conv2D and ReLU) and diverse usage scenarios (i.e., operator parameter configurations), into unified IRs, requiring comprehensive coverage of library-specific semantics.
(2) \textit{Optimization-stage awareness:}
The high-level optimization stage performs complex graph-level transformations, making it difficult to generate computational graphs that effectively trigger hardware-independent optimizations.
(3) \textit{Hardware-specific validation:}
The low-level optimization stage applies hardware-specific optimizations and generates executable code for target platforms, requiring tests that incorporate features capable of triggering the corresponding hardware-specific optimization logic.

\textbf{Our work:}
To address these challenges, we propose a unified data-driven AI compiler testing framework integrating three complementary components: a migration-based test generator for the model loading stage, a synthesis-based test generator for the high-level optimization stage, and a mutation-based test generator for the low-level optimization stage.
Specifically, the framework comprises OPERA, a migration-based testing technique driven by operator tests for AI frameworks.
It then incorporates OATest, an optimization-aware test synthesis technique guided by optimization-related tests crafted by compiler developers. 
Finally, it introduces HARMONY, a mutation-based technique that leverages documentation of hardware-specific optimizations and knowledge from large language models (LLM).
Together, these components form a unified approach to enhancing the reliability of AI compilers across the entire compilation pipeline.
Our key contributions are as follows:
\begin{itemize}
    \item \textbf{OPERA:} We propose a migration-based approach that leverages existing AI library tests to maximize coverage of diverse operator conversion logic in the model loading stage, uncovering 170 bugs.
    \item \textbf{OATest:} We design an optimization-aware test generation technique that extracts semantic patterns from documented tests and synthesizes optimization-sensitive tests by integrating them with diverse contexts, revealing 56 bugs in high-level optimization stages.
    \item \textbf{HARMONY:} We develop a mutation-based test generation technique that first generates diverse seeds to cover different computation scenarios, and then mutates them to produce hardware-optimization-oriented tests, guided by optimization patterns extracted from documentation and LLMs, detecting 40 bugs in low-level optimization stages.
\end{itemize}

%% file: 3-related.tex
\section{Background and related work}
\label{related}

AI compilers typically follow a three-stage pipeline comprising model loading, high-level optimization~\cite{li2020deep}, and low-level optimization.
In the model loading stage, AI models built with AI libraries such as PyTorch~\cite{pytorch} or TensorFlow~\cite{tensorflow} are converted into a unified high-level intermediate representation (IR).
During the high-level optimization stage, the compiler applies hardware-independent transformations, including operator fusion, layout inference, and constant folding.
Finally, in the low-level optimization stage, the compiler performs hardware-specific optimizations (e.g., memory latency hiding and parallelization) on the low-level IR to generate executable code for target hardware~\cite{guo2019empirical}.

Existing AI compiler testing techniques can be classified into three categories according to the test formats they generate: model generators, high-level IR generators, and low-level IR generators.
Specifically, model generators (e.g., NNSmith~\cite{NNSmith}, WhiteFox~\cite{whitefox}, Polyjuice~\cite{polyjuice}) produce AI models that, while capable of exercising the full compilation stack, in practice focus mainly on optimization stages rather than fully covering the model loading stage.
High-level IR generators (e.g., HirGen~\cite{HirGen}, GenCoG~\cite{gencog}) target high-level optimizations by generating diverse computational graphs to trigger transformations such as operator fusion and layout inference.
Low-level IR generators (e.g., Tzer~\cite{tzer}) focus on hardware-specific optimizations by mutating low-level IR to activate backend transformations such as memory latency hiding and instruction scheduling.

Despite these advances, existing methods primarily aim to generate diverse yet valid tests, without fully covering the core logic of each compilation stage. Consequently, they are unable to achieve efficient, stage-specific test generation for AI compilers.

%% file: 2-approach.tex
\section{Finished Work}
This section provides evidence supporting our contributions. Specifically, we present the results corresponding to each contribution. As discussed in Section~\ref{intro}, our research focuses on generating effective tests to fuzz each stage of AI compilers.

\textbf{OPERA: Migration-Based Model Loading Testing}.
OPERA introduces a migration-based technique for the model loading stage. 
Since AI operators are implemented and validated by AI libraries, their library tests already exercise the core semantics of individual operators. Meanwhile, the model loading stage of AI compilers converts operators from diverse frameworks into unified IRs, a process that closely parallels the operator behavior verified by these library tests.
Building on this insight, OPERA first instruments all APIs for constructing AI models in the source code of AI libraries to extract operator instances by executing library tests.
It then encapsulates the extracted operator instance into a single-operator model through template-based generation. To address the significant cost introduced by the large number of migrated tests and the need for frequent reruns due to rapidly evolving AI libraries and compilers, OPERA employs a diversity-aware prioritization algorithm that clusters operators by semantics to maximize coverage while eliminating redundancy.
Evaluated on eight frontends of TVM~\cite{TVM}, TensorRT~\cite{TensorRT}, and OpenVINO~\cite{OpenVINO}, OPERA uncovered 170 new bugs, 90 of which were confirmed by developers.
The root causes of these bugs include Tensor Shape Problem, Type Problem, Incorrect Code Logic, Incorrect Exception Handling, Incompatibility, Incorrect Assignment, Incorrect Numerical Computation, Concurrency, and Typo, covering all root cause categories summarized on historical bugs in the model loading stage~\cite{DLCStudy}.
The prioritization strategy improved testing efficiency by  11.9\%$\sim$47.4\% on average compared to general test prioritization strategies, validating test migration as a powerful strategy for efficient migration. This work was published in ICSE 2025~\cite{opera}.

\textbf{OATest: Synthesis-Based High-level Optimization Testing.}
OATest introduces a synthesis-based testing technique targeting high-level optimizations of AI compilers.
Since optimizations on high-level IRs (i.e., computational graph) are context-sensitive, the same optimization pattern may lead to different transformation paths when placed in different contexts. 
Leveraging this insight, we proposed OATest, a novel optimization-aware computational graphs generation technique, for testing high-level optimizations.
Specifically, OATest first instruments the source code of AI compilers to extract optimization patterns by executing all optimization tests crafted by AI compiler developers.
Since different optimizations operate at varying levels of granularity, extracting optimization-aware patterns from the documented tests involves two steps: (1) identifying the computational
graph and its corresponding optimization, and (2) deriving optimization-aware patterns from the computational graph based on the granularity of the optimization.
The extracted patterns are then combined with diverse contexts drawn from seed graphs to generate optimization-aware tests.
Here, OATest randomly selects both a pattern and a seed graph, and inserts the pattern into the seed graph at a randomly chosen synthesis point. These random choices enhance the structural diversity of the generated computational graphs.
However, naively combining patterns and contexts can result in invalid tests due to undefined operator inputs within the inserted patterns. To guarantee test validity, we design two fixing strategies: (1) reusing compatible inputs and outputs from existing nodes, and (2) creating new nodes with compatible inputs and outputs to establish correct and effective connections between patterns and contexts.
Evaluated on two popular AI compilers (i.e., TVM and ONNXRuntime), OATest discovered 56 previously unknown bugs, 42 of which were confirmed with 24 fixed.
Compared to NNSmith~\cite{NNSmith} and WhiteFox~\cite{whitefox}, it achieved 60.2\% higher branch coverage and 66.98\% higher line coverage. This work has been accepted to ICSE 2026~\cite{oatest}.

\textbf{HARMONY: Mutation-Based Low-level Optimization Testing.}
HARMONY presents a mutation-based framework for testing hardware-specific optimizations. 
Since optimizations on low-level IRs (i.e., hardware-specific optimizations) are difficult to trigger due to complex constraints and deep invocation chains, we directly use diversified low-level IRs as seeds. Lightweight mutations are then applied to preserve validity while effectively generating optimization-sensitive test cases.
Specifically, HARMONY first leverages operator constraints collected from multiple knowledge sources to guide LLMs in generating valid and diverse low-level IRs as seeds.
These cross-checked knowledge sources mitigate the risk of invalid tests caused by outdated or incorrect constraints, which are common in practice due to the rapid evolution of AI compilers and lagging documentation updates.
Rather than generating tests from scratch, HARMONY applies lightweight mutations to existing seeds, thereby reducing the risk of invalid outputs caused by LLM hallucinations. Optimization patterns are extracted from documentation and code examples using an analysis LLM. A code LLM then mutates the IR seeds according to these patterns, producing semantically valid tests that are likely to trigger targeted low-level optimizations, such as memory latency hiding and intrinsic mapping.
Evaluated on the popular AI compiler (i.e., TVM), HARMONY uncovered 40 previously unknown bugs, 26 of which were confirmed and fixed. Compared to state-of-the-art fuzzers such as NNSmith~\cite{NNSmith}, WhiteFox~\cite{whitefox}, and Tzer~\cite{tzer}, HARMONY achieves substantially higher coverage at the low-level optimization stage, demonstrating its effectiveness in detecting hardware-level optimization vulnerabilities.
This work is ongoing, with plans to further validate the approach across additional AI compilers.

%% file: 4-conclusion.tex
\section{Conclusion and Future Work}
This work presents a unified framework for testing AI compilers. OPERA improves coverage of the model loading stage by migrating operator tests for AI libraries. OATest targets context-sensitive high-level optimizations by synthesizing optimization-aware computational graphs. HARMONY addresses hardware-specific low-level optimizations through mutation-based IR generation guided by multiple knowledge sources. Together, these methods systematically detect bugs across the AI compilation pipeline, uncovering 266 previously unknown bugs.

Future work will extend this framework to emerging AI compilers, such as Triton, Tilelang, MLC-LLM, and TensorRT-LLM. By adapting our framework to these new and evolving AI compilers, we aim to further advance reliable, stage-specific testing across the AI compiler landscape. In parallel, automatic bug localization and repair techniques are also part of our future research, aiming to reduce developers’ effort in debugging and maintenance. I plan to continue investigating these topics during my postdoctoral research and in my future role as a faculty member and researcher.

\begin{acks}
    Qingchao Shen is a third-year Ph.D. student, out of an expected four years, under the supervision of Prof. Junjie Chen. 
    This work is supported by National Key Research and Development Program of China (No. 2024YFB4506300), and National Natural Science Foundation of China (Grant No. 62322208).
\end{acks}

%% file: 0-main.bbl

\begin{thebibliography}{27}

%%% the Web address.

\ifx \showCODEN    \undefined \def \showCODEN     #1{\unskip}     \fi
\ifx \showDOI      \undefined \def \showDOI       #1{#1}\fi
\ifx \showISBNx    \undefined \def \showISBNx     #1{\unskip}     \fi
\ifx \showISBNxiii \undefined \def \showISBNxiii  #1{\unskip}     \fi
\ifx \showISSN     \undefined \def \showISSN      #1{\unskip}     \fi
\ifx \showLCCN     \undefined \def \showLCCN      #1{\unskip}     \fi
\ifx \shownote     \undefined \def \shownote      #1{#1}          \fi
\ifx \showarticletitle \undefined \def \showarticletitle #1{#1}   \fi
\ifx \showURL      \undefined \def \showURL       {\relax}        \fi
% invisible to TeX
\providecommand\bibfield[2]{#2}
\providecommand\bibinfo[2]{#2}
\providecommand\natexlab[1]{#1}
\providecommand\showeprint[2][]{arXiv:#2}

\bibitem[Ope(2023)]%
        {OpenVINO}
 \bibinfo{year}{Accessed: 2023}\natexlab{}.
\newblock \bibinfo{title}{Intel OpenVINO}.
\newblock
\newblock
\newblock
\shownote{\url{https://docs.openvino.ai/2022.3/home.html}}.


\bibitem[ker(2025)]%
        {keras}
 \bibinfo{year}{Accessed: 2025}\natexlab{}.
\newblock \bibinfo{title}{Keras}.
\newblock
\newblock
\newblock
\shownote{\url{https://keras.io/}}.


\bibitem[gpu(2025)]%
        {gpu}
 \bibinfo{year}{Accessed: 2025}\natexlab{}.
\newblock \bibinfo{title}{NVIDIA GPU}.
\newblock
\newblock
\newblock
\shownote{\url{https://www.nvidia.com}}.


\bibitem[Ten(2025)]%
        {TensorRT}
 \bibinfo{year}{Accessed: 2025}\natexlab{}.
\newblock \bibinfo{title}{NVIDIA TensorRT}.
\newblock
\newblock
\newblock
\shownote{\url{https://developer.nvidia.com/tensorrt}}.


\bibitem[onn(2025)]%
        {onnx}
 \bibinfo{year}{Accessed: 2025}\natexlab{}.
\newblock \bibinfo{title}{ONNX}.
\newblock
\newblock
\newblock
\shownote{\url{https://onnx.ai/onnx/}}.


\bibitem[pyt(2025)]%
        {pytorch}
 \bibinfo{year}{Accessed: 2025}\natexlab{}.
\newblock \bibinfo{title}{PyTorch}.
\newblock
\newblock
\newblock
\shownote{\url{https://pytorch.org/}}.


\bibitem[ten(2025)]%
        {tensorflow}
 \bibinfo{year}{Accessed: 2025}\natexlab{}.
\newblock \bibinfo{title}{TensorFlow}.
\newblock
\newblock
\newblock
\shownote{\url{https://www.tensorflow.org/}}.


\bibitem[TVM(2025)]%
        {TVMFuzz}
 \bibinfo{year}{Accessed: 2025}\natexlab{}.
\newblock \bibinfo{title}{TVMFuzz}.
\newblock
\newblock
\newblock
\shownote{\url{https://github.com/dpankratz/TVMFuzz}}.


\bibitem[Chen et~al\mbox{.}(2018)]%
        {TVM}
\bibfield{author}{\bibinfo{person}{Tianqi Chen}, \bibinfo{person}{Thierry Moreau}, \bibinfo{person}{Ziheng Jiang}, \bibinfo{person}{Lianmin Zheng}, \bibinfo{person}{Eddie Yan}, \bibinfo{person}{Haichen Shen}, \bibinfo{person}{Meghan Cowan}, \bibinfo{person}{Leyuan Wang}, \bibinfo{person}{Yuwei Hu}, \bibinfo{person}{Luis Ceze}, {et~al\mbox{.}}} \bibinfo{year}{2018}\natexlab{}.
\newblock \showarticletitle{TVM: An automated End-to-End optimizing compiler for deep learning}. In \bibinfo{booktitle}{\emph{13th USENIX Symposium on Operating Systems Design and Implementation (OSDI 18)}}. \bibinfo{pages}{578--594}.
\newblock


\bibitem[Guo et~al\mbox{.}(2019)]%
        {guo2019empirical}
\bibfield{author}{\bibinfo{person}{Qianyu Guo}, \bibinfo{person}{Sen Chen}, \bibinfo{person}{Xiaofei Xie}, \bibinfo{person}{Lei Ma}, \bibinfo{person}{Qiang Hu}, \bibinfo{person}{Hongtao Liu}, \bibinfo{person}{Yang Liu}, \bibinfo{person}{Jianjun Zhao}, {and} \bibinfo{person}{Xiaohong Li}.} \bibinfo{year}{2019}\natexlab{}.
\newblock \showarticletitle{An empirical study towards characterizing deep learning development and deployment across different frameworks and platforms}. In \bibinfo{booktitle}{\emph{Proceedings of 34th IEEE/ACM International Conference on Automated Software Engineering}}. IEEE, \bibinfo{pages}{810--822}.
\newblock


\bibitem[Huang et~al\mbox{.}(2025)]%
        {huang2025false}
\bibfield{author}{\bibinfo{person}{Lili Huang}, \bibinfo{person}{Qingchao Shen}, \bibinfo{person}{Dong Wang}, \bibinfo{person}{Yunping Wu}, \bibinfo{person}{Meng Wang}, {and} \bibinfo{person}{Junjie Chen}.} \bibinfo{year}{2025}\natexlab{}.
\newblock \showarticletitle{False-Positive Bug Reports in Deep Learning Compilers: Stages, Root Causes, and Mitigation}.
\newblock \bibinfo{journal}{\emph{ACM Transactions on Software Engineering and Methodology}} (\bibinfo{year}{2025}).
\newblock


\bibitem[Jouppi et~al\mbox{.}(2017)]%
        {tpu}
\bibfield{author}{\bibinfo{person}{Norman~P Jouppi}, \bibinfo{person}{Cliff Young}, \bibinfo{person}{Nishant Patil}, \bibinfo{person}{David Patterson}, \bibinfo{person}{Gaurav Agrawal}, \bibinfo{person}{Raminder Bajwa}, \bibinfo{person}{Sarah Bates}, \bibinfo{person}{Suresh Bhatia}, \bibinfo{person}{Nan Boden}, \bibinfo{person}{Al Borchers}, {et~al\mbox{.}}} \bibinfo{year}{2017}\natexlab{}.
\newblock \showarticletitle{In-datacenter performance analysis of a tensor processing unit}. In \bibinfo{booktitle}{\emph{Proceedings of the 44th annual international symposium on computer architecture}}. \bibinfo{pages}{1--12}.
\newblock


\bibitem[Li et~al\mbox{.}(2020)]%
        {li2020deep}
\bibfield{author}{\bibinfo{person}{Mingzhen Li}, \bibinfo{person}{Yi Liu}, \bibinfo{person}{Xiaoyan Liu}, \bibinfo{person}{Qingxiao Sun}, \bibinfo{person}{Xin You}, \bibinfo{person}{Hailong Yang}, \bibinfo{person}{Zhongzhi Luan}, \bibinfo{person}{Lin Gan}, \bibinfo{person}{Guangwen Yang}, {and} \bibinfo{person}{Depei Qian}.} \bibinfo{year}{2020}\natexlab{}.
\newblock \showarticletitle{The deep learning compiler: A comprehensive survey}.
\newblock \bibinfo{journal}{\emph{IEEE Transactions on Parallel and Distributed Systems}} \bibinfo{volume}{32}, \bibinfo{number}{3} (\bibinfo{year}{2020}), \bibinfo{pages}{708--727}.
\newblock


\bibitem[Liu et~al\mbox{.}(2023)]%
        {NNSmith}
\bibfield{author}{\bibinfo{person}{Jiawei Liu}, \bibinfo{person}{Jinkun Lin}, \bibinfo{person}{Fabian Ruffy}, \bibinfo{person}{Cheng Tan}, \bibinfo{person}{Jinyang Li}, \bibinfo{person}{Aurojit Panda}, {and} \bibinfo{person}{Lingming Zhang}.} \bibinfo{year}{2023}\natexlab{}.
\newblock \showarticletitle{NNSmith: Generating Diverse and Valid Test Cases for Deep Learning Compilers}. In \bibinfo{booktitle}{\emph{Proceedings of the 28th ACM International Conference on Architectural Support for Programming Languages and Operating Systems, Volume 2}} (Vancouver, BC, Canada) \emph{(\bibinfo{series}{ASPLOS 2023})}. \bibinfo{publisher}{Association for Computing Machinery}, \bibinfo{address}{New York, NY, USA}, \bibinfo{pages}{530–543}.
\newblock
\showISBNx{9781450399166}
\urldef\tempurl%
\url{https://doi.org/10.1145/3575693.3575707}
\showDOI{\tempurl}


\bibitem[Liu et~al\mbox{.}(2022)]%
        {tzer}
\bibfield{author}{\bibinfo{person}{Jiawei Liu}, \bibinfo{person}{Yuxiang Wei}, \bibinfo{person}{Sen Yang}, \bibinfo{person}{Yinlin Deng}, {and} \bibinfo{person}{Lingming Zhang}.} \bibinfo{year}{2022}\natexlab{}.
\newblock \showarticletitle{Coverage-Guided Tensor Compiler Fuzzing with Joint IR-Pass Mutation}.
\newblock \bibinfo{journal}{\emph{Proc. ACM Program. Lang.}} \bibinfo{volume}{6}, \bibinfo{number}{OOPSLA1}, Article \bibinfo{articleno}{73} (\bibinfo{date}{Apr} \bibinfo{year}{2022}), \bibinfo{numpages}{26}~pages.
\newblock
\urldef\tempurl%
\url{https://doi.org/10.1145/3527317}
\showDOI{\tempurl}


\bibitem[Ma et~al\mbox{.}(2025)]%
        {ma2025bounded}
\bibfield{author}{\bibinfo{person}{Haoyang Ma}, \bibinfo{person}{Alastair~F Donaldson}, \bibinfo{person}{Qingchao Shen}, \bibinfo{person}{Yongqiang Tian}, \bibinfo{person}{Junjie Chen}, {and} \bibinfo{person}{Shing-Chi Cheung}.} \bibinfo{year}{2025}\natexlab{}.
\newblock \showarticletitle{Bounded Exhaustive Random Program Generation for Testing Solidity Compilers and Analyzers}.
\newblock \bibinfo{journal}{\emph{arXiv preprint arXiv:2503.20332}} (\bibinfo{year}{2025}).
\newblock


\bibitem[Ma et~al\mbox{.}(2023)]%
        {HirGen}
\bibfield{author}{\bibinfo{person}{Haoyang Ma}, \bibinfo{person}{Qingchao Shen}, \bibinfo{person}{Yongqiang Tian}, \bibinfo{person}{Junjie Chen}, {and} \bibinfo{person}{Shing-Chi Cheung}.} \bibinfo{year}{2023}\natexlab{}.
\newblock \showarticletitle{Fuzzing Deep Learning Compilers with HirGen}. In \bibinfo{booktitle}{\emph{Proceedings of the 32nd ACM SIGSOFT International Symposium on Software Testing and Analysis}} (Seattle, WA, USA) \emph{(\bibinfo{series}{ISSTA 2023})}. \bibinfo{publisher}{Association for Computing Machinery}, \bibinfo{address}{New York, NY, USA}, \bibinfo{pages}{248–260}.
\newblock
\showISBNx{9798400702211}
\urldef\tempurl%
\url{https://doi.org/10.1145/3597926.3598053}
\showDOI{\tempurl}


\bibitem[Shen et~al\mbox{.}(2022)]%
        {qaqa}
\bibfield{author}{\bibinfo{person}{Qingchao Shen}, \bibinfo{person}{Junjie Chen}, \bibinfo{person}{Jie~M Zhang}, \bibinfo{person}{Haoyu Wang}, \bibinfo{person}{Shuang Liu}, {and} \bibinfo{person}{Menghan Tian}.} \bibinfo{year}{2022}\natexlab{}.
\newblock \showarticletitle{Natural test generation for precise testing of question answering software}. In \bibinfo{booktitle}{\emph{Proceedings of the 37th IEEE/ACM International Conference on Automated Software Engineering}}. \bibinfo{pages}{1--12}.
\newblock


\bibitem[Shen et~al\mbox{.}(2021)]%
        {DLCStudy}
\bibfield{author}{\bibinfo{person}{Qingchao Shen}, \bibinfo{person}{Haoyang Ma}, \bibinfo{person}{Junjie Chen}, \bibinfo{person}{Yongqiang Tian}, \bibinfo{person}{Shing-Chi Cheung}, {and} \bibinfo{person}{Xiang Chen}.} \bibinfo{year}{2021}\natexlab{}.
\newblock \showarticletitle{A Comprehensive Study of Deep Learning Compiler Bugs} \emph{(\bibinfo{series}{ESEC/FSE 2021})}. \bibinfo{publisher}{Association for Computing Machinery}, \bibinfo{address}{New York, NY, USA}, \bibinfo{pages}{968–980}.
\newblock
\showISBNx{9781450385626}
\urldef\tempurl%
\url{https://doi.org/10.1145/3468264.3468591}
\showDOI{\tempurl}


\bibitem[Shen et~al\mbox{.}(2024)]%
        {dlc_evolution}
\bibfield{author}{\bibinfo{person}{Qingchao Shen}, \bibinfo{person}{Jiashuo Tian}, \bibinfo{person}{Junjie Chen}, \bibinfo{person}{Xiang Chen}, \bibinfo{person}{Qingyan Chen}, {and} \bibinfo{person}{Zan Wang}.} \bibinfo{year}{2024}\natexlab{}.
\newblock \showarticletitle{Toward Understanding the Current Status and Evolution of Deep Learning Compiler Bugs}.
\newblock \bibinfo{journal}{\emph{Journal of Software}} \bibinfo{volume}{36}, \bibinfo{number}{7} (\bibinfo{year}{2024}), \bibinfo{pages}{3022--3040}.
\newblock


\bibitem[Shen et~al\mbox{.}(2025a)]%
        {opera}
\bibfield{author}{\bibinfo{person}{Qingchao Shen}, \bibinfo{person}{Yongqiang Tian}, \bibinfo{person}{Haoyang Ma}, \bibinfo{person}{Junjie Chen}, \bibinfo{person}{Lili Huang}, \bibinfo{person}{Ruifeng Fu}, \bibinfo{person}{Shing-Chi Cheung}, {and} \bibinfo{person}{Zan Wang}.} \bibinfo{year}{2025}\natexlab{a}.
\newblock \showarticletitle{A Tale of Two DL Cities: When Library Tests Meet Compiler}. In \bibinfo{booktitle}{\emph{2025 IEEE/ACM 47th International Conference on Software Engineering (ICSE)}}. IEEE, \bibinfo{pages}{2201--2212}.
\newblock


\bibitem[Shen et~al\mbox{.}(2025b)]%
        {oatest}
\bibfield{author}{\bibinfo{person}{Qingchao Shen}, \bibinfo{person}{Zan Wang}, \bibinfo{person}{Haoyang Ma}, \bibinfo{person}{Yongqiang Tian}, \bibinfo{person}{Lili Huang}, \bibinfo{person}{Zibo Xiao}, \bibinfo{person}{Junjie Chen}, {and} \bibinfo{person}{Shing-Chi Cheung}.} \bibinfo{year}{2025}\natexlab{b}.
\newblock \showarticletitle{Optimization-Aware Test Generation for Deep Learning Compilers}.
\newblock \bibinfo{journal}{\emph{arXiv preprint arXiv:2511.18918}} (\bibinfo{year}{2025}).
\newblock


\bibitem[Suo et~al\mbox{.}(2025)]%
        {desil}
\bibfield{author}{\bibinfo{person}{Chenyao Suo}, \bibinfo{person}{Jianrong Wang}, \bibinfo{person}{Yongjia Wang}, \bibinfo{person}{Jiajun Jiang}, \bibinfo{person}{Qingchao Shen}, {and} \bibinfo{person}{Junjie Chen}.} \bibinfo{year}{2025}\natexlab{}.
\newblock \showarticletitle{DESIL: Detecting Silent Bugs in MLIR Compiler Infrastructure}.
\newblock \bibinfo{journal}{\emph{Proceedings of the ACM on Programming Languages}} \bibinfo{volume}{9}, \bibinfo{number}{OOPSLA2} (\bibinfo{year}{2025}), \bibinfo{pages}{3093--3119}.
\newblock


\bibitem[Wang et~al\mbox{.}(2023a)]%
        {mlirsmith}
\bibfield{author}{\bibinfo{person}{Haoyu Wang}, \bibinfo{person}{Junjie Chen}, \bibinfo{person}{Chuyue Xie}, \bibinfo{person}{Shuang Liu}, \bibinfo{person}{Zan Wang}, \bibinfo{person}{Qingchao Shen}, {and} \bibinfo{person}{Yingquan Zhao}.} \bibinfo{year}{2023}\natexlab{a}.
\newblock \showarticletitle{Mlirsmith: Random program generation for fuzzing mlir compiler infrastructure}. In \bibinfo{booktitle}{\emph{2023 38th IEEE/ACM International Conference on Automated Software Engineering (ASE)}}. IEEE, \bibinfo{pages}{1555--1566}.
\newblock


\bibitem[Wang et~al\mbox{.}(2023b)]%
        {gencog}
\bibfield{author}{\bibinfo{person}{Zihan Wang}, \bibinfo{person}{Pengbo Nie}, \bibinfo{person}{Xinyuan Miao}, \bibinfo{person}{Yuting Chen}, \bibinfo{person}{Chengcheng Wan}, \bibinfo{person}{Lei Bu}, {and} \bibinfo{person}{Jianjun Zhao}.} \bibinfo{year}{2023}\natexlab{b}.
\newblock \showarticletitle{GenCoG: A DSL-Based Approach to Generating Computation Graphs for TVM Testing}. In \bibinfo{booktitle}{\emph{Proceedings of the 32nd ACM SIGSOFT International Symposium on Software Testing and Analysis}}. \bibinfo{pages}{904--916}.
\newblock


\bibitem[Yang et~al\mbox{.}(2023)]%
        {whitefox}
\bibfield{author}{\bibinfo{person}{Chenyuan Yang}, \bibinfo{person}{Yinlin Deng}, \bibinfo{person}{Runyu Lu}, \bibinfo{person}{Jiayi Yao}, \bibinfo{person}{Jiawei Liu}, \bibinfo{person}{Reyhaneh Jabbarvand}, {and} \bibinfo{person}{Lingming Zhang}.} \bibinfo{year}{2023}\natexlab{}.
\newblock \showarticletitle{White-box compiler fuzzing empowered by large language models}.
\newblock \bibinfo{journal}{\emph{arXiv preprint arXiv:2310.15991}} (\bibinfo{year}{2023}).
\newblock


\bibitem[Zhou et~al\mbox{.}(2024)]%
        {polyjuice}
\bibfield{author}{\bibinfo{person}{Chijin Zhou}, \bibinfo{person}{Bingzhou Qian}, \bibinfo{person}{Gwihwan Go}, \bibinfo{person}{Quan Zhang}, \bibinfo{person}{Shanshan Li}, {and} \bibinfo{person}{Yu Jiang}.} \bibinfo{year}{2024}\natexlab{}.
\newblock \showarticletitle{Polyjuice: Detecting mis-compilation bugs in tensor compilers with equality saturation based rewriting}.
\newblock \bibinfo{journal}{\emph{Proceedings of the ACM on Programming Languages}} \bibinfo{volume}{8}, \bibinfo{number}{OOPSLA2} (\bibinfo{year}{2024}), \bibinfo{pages}{1309--1335}.
\newblock


\end{thebibliography}
